\pdfoutput=1   
\def\lcversion{}                        

\documentclass[letterpaper, conference, onecolumn]{ieeeconf}
\def\MODE{3}

\usepackage{cite}                   
\usepackage{math}                   
\usepackage[utf8]{inputenc}			
\usepackage[english]{babel}			
\usepackage[hidelinks]{hyperref}    

\if\MODE3\else
\usepackage{amsthm}
\newtheorem{thm}{Theorem}
\newtheorem{problem}[thm]{Problem}
\newtheorem{lem}[thm]{Lemma}
\newtheorem{prop}[thm]{Proposition}
\newtheorem{rem}[thm]{Remark}
\newtheorem{defn}[thm]{Definition}
\newtheorem{cor}[thm]{Corollary}
\newtheorem{assumption}[thm]{Assumption}
\def\qed{\rule[0pt]{5pt}{5pt}\par\medskip}
\renewcommand{\qedhere}{\hfill ~\qed}
\renewenvironment{proof}{{\noindent\bf Proof.}}{\qedhere}
\fi

\usepackage{bbm}                    
\usepackage{textcomp}				
\usepackage{graphicx}
\graphicspath{{./graphics/}}

\usepackage{tikz}
\usepackage{amsmath}
\usetikzlibrary{calc,positioning}
\usepackage{amssymb}
\if\MODE3
\fi

\usepackage{ntheorem}
\usepackage[capitalise,nameinlink]{cleveref}
\usetikzlibrary{shapes.geometric}

\newtheorem{thm}{Theorem}

\crefname{thm}{Theorem}{Theorems}
\crefname{problem}{Problem}{Problems}
\crefname{lem}{Lemma}{Lemmas}
\crefname{prop}{Proposition}{Propositions}
\crefname{rem}{Remark}{Remarks}
\crefname{defn}{Definition}{Definitions}
\crefname{cor}{Corollary}{Corollaries}
\crefname{assumption}{Assumption}{Assumptions}

\begin{document}
\title{Sliding Mode Roll Control of Active Suspension Electric Vehicles}

\if\MODE1
\author{Mruganka Kashyap}
\note{Submitted to ACC 2025}
\fi

\if\MODE2
\author{Mruganka Kashyap$^{1}$}
\note{Submitted to ACC 2025}
\fi

\if\MODE3
\def\BibTeX{{\rm B\kern-.05em{\sc i\kern-.025em b}\kern-.08em
    T\kern-.1667em\lower.7ex\hbox{E}\kern-.125emX}}
\markboth{\journalname, VOL. XX, NO. XX, XXXX 2023}
{Kashyap \MakeLowercase{\textit{et al.}}: Preparation of Papers for IEEE Control Systems Letters (2023)}

\author{
	Mruganka Kashyap$^{1}$ \IEEEmembership{Member, IEEE}\\
	Submitted to the 13th IFAC Symposium on Nonlinear Control Systems
	\thanks{M.~Kashyap is a Senior Controls Engineer at ASML. (e-mail: mruganka.kashyap@asml.com)}
}
\fi

\maketitle
\thispagestyle{empty}

\if\MODE3
\footnotetext[1]{M.~Kashyap is a Senior Controls Engineer with the Light Control Group at Cymer, LLC, an ASML company, San Diego, CA 92069, USA. (e-mail: mruganka.kashyap@asml.com).
	
 }
\fi


\begin{abstract}
	Vehicle roll control has been a well studied problem. One of the ubiquitous methods to mitigate vehicle rollover in the automobile industry is via a mechanical anti-roll bar. However with the advent of electric vehicles, rollover mitigation can be pursued using electric actuation. In this work, we study a roll control algorithm using sliding mode control for active suspension vehicles, where the actuation for the roll control signal is generated by electric motors independently at the four corners of the vehicle. This technology precludes the need for any mechanical actuation which is often slower as well as any anti-roll bar to mitigate vehicle rollover situations. We provide an implementation of the proposed algorithm and conduct numerical experiments to validate the functionality and effectiveness. Specifically, we perform Slalom and J-turn maneuvering tests on an active suspension electric vehicle with sliding model roll control and it is shown to mitigate rollover by atleast 50$\%$ compared to passive suspension vehicles, while simultaneously maintaining rider comfort.                                        
\end{abstract}


\section{Introduction}\label{sec:intro}

The
development and integration of active suspension systems for electric vehicles has been the subject of extensive research in the automobile industry. The rationale for better suspensions in automobiles stems from the need to achieve a greater degree of ride quality and electric vehicle handling by keeping the chassis of the electric vehicle parallel to the road when turning corners, and reducing the impact of disturbances from the road surface on the vehicle. Vehicle suspensions can be categorized into three broad categories: passive, semi-active, and active suspensions. Passive suspension systems constitute large springs at the automobile wheels that dampen the vertical movements caused by the road surface on the vehicle body. The parameters of passive suspension, specifically the spring constant and damping coefficient, are fixed to ensure a certain degree of balance between reducing road surface transmitted disturbances and enhancing road holding. The former enhances ride quality while the later ensures dynamic stability of the vehicle.

In this paper, we consider a electric vehicle body equipped with active suspensions at each of the four corners, where a motor provides a linear, vertical force.  We also consider that pitch, heave, and roll are simultaneously controlled via these vertical forces. This assumption is critical to the optimal distribution of the roll control commands to the electric vehicle corners. However, the focus of this paper would on roll control of the vehicle that is synonymous with vehicle stability. 

In \cref{sec:Decen} we provide a detailed review of the roll dynamics model for active suspension vehicles. In \cref{sec:main} we provide the sliding mode roll control algorithm for active suspension electric vehicles and we present the corresponding implementation in \cref{subsec:implementation}. In \cref{sec:experiments} we establish the validity of the proposed roll control algorithm via MATLAB simulations on an active suspension electric vehicle. We conclude in \cref{sec:conclufuture}.

\section{Roll model}\label{sec:Decen}
\subsection{Roll dynamics model}
\label{subsec:roll_model}
We consider a six degrees of freedom (DoF) roll dynamics model in this section. In active rollover situations, the front of the vehicle undergoes rollover prior to the vehicle rear. We incorporate this assumption while designing our dynamic 6-DoF roll model. The 6 DoF considered include heave of the sprung mass, roll angle of the sprung mass, and vertical translations of the left and right sprung masses. The consolidated equations of motion are:
\begin{align}\label{eq:roll}
   \tilde{I}_{xx} \ddot{\phi}&= m_s a_y h_{\phi} \textsf{cos}\phi +m_s g h_{\phi} \textsf{sin}\phi
    -0.5 k_f l_s^2 \textsf{sin}\phi -0.5 b_f l_s^2 \dot{\phi}\textsf{cos}\phi
    -0.5 k_r l_s^2 \textsf{sin}\phi -0.5 b_r l_s^2 \dot{\phi}\textsf{cos}\phi\\
    &-0.5 k_f l_s \left(z_{u,fl}-z_{u,fr}\right) -0.5 b_f l_s \left(\dot{z}_{u,fl}-\dot{z}_{u,fr}\right)
    -0.5 k_r l_s \left(z_{u,rl}-z_{u,rr}\right) -0.5 b_r l_s \left(\dot{z}_{u,rl}-\dot{z}_{u,rr}\right),\nonumber\\ 
    m_{u,ij} \ddot{z}_{u,ij} &= k_i \left(z_{s}-z_{u,ij}+0.5\;(-1)^{\textsf{sign}(\phi)} \textsf{sin}\phi\right)
    + b_i \left(\dot{z}_{s}-\dot{z}_{u,ij}+0.5\;(-1)^{\textsf{sign}(\phi)} \dot{\phi}\textsf{cos}\phi\right)
    -k_t \left(z_{u,ij}-z_{\textsf{road}}\right),
\end{align}
where $\tilde{I}_{xx} = I_{xx}+m_s h_{\phi}^2$, $i\in\{\textsf{front}\; (f), \textsf{rear}\; (r)\}$, $j\in\{\textsf{left}\; (l), \textsf{right}\; (r)\}$, $m_{u,ij}$ is the unsprung mass for the four corners, $m_s$ is the sprung mass, $a_y$ is lateral acceleration at the center of gravity, $\phi$ is the roll angle, $z_{u,ij}$ is the relative displacement of the unsprung mass from the quarter-car model, $h_{\phi}$ is the distance of the roll center from the ground, $g$ is the acceleration due to gravity, $k_f$ and $k_r$ are the stiffness constants for front and rear suspensions, $b_f$ and $b_r$ are the corresponding damping constants, $l_s$ is the distance between the left and right suspensions, $k_t$ is the stiffness constant for the tires. We make the assumption that $l_s$ is equal for both front and right suspensions.

\section{Sliding mode roll control}\label{sec:main}
Here we provide the sliding mode roll control algorithm that is implemented by four electrical actuators (as part of the unsprung masses) at the vehicle corners via the suspension connecting the sprung and unsprung masses. While there exist attempts to leverage the concept of sliding mode control to develop anti-roll systems, the previous works utilize anti-roll bars or other mechanical actuation \cite{al2002sliding,konieczny2020active,chu2015smooth,zhang2018sliding}. The proposed method in the paper differs in that it is based on a 6 DoF roll model, leverages electric actuation at the unsprung masses, and is distributed in implementation.  
\begin{thm}\label{thm:sliding_mode}
    Consider the roll control model defined in \cref{subsec:roll_model}. Define a feedback roll control signal $u_{\phi}$ that ensures electric vehicle stability by reducing the roll angle $\left(\phi\right)$, while simultaneously maintaining rider comfort by minimizing the roll rate $(\dot{\phi})$. Then the roll control signal is given by
    \begin{multline}
        u_{\phi} = - \tilde{I}_{xx}\frac{\eta}{\psi} \phi - \tilde{I}_{xx}\left(\eta+\frac{1}{\psi}\right)\dot{\phi} -m_s a_y h_{\phi} \textsf{cos}\phi -m_s g h_{\phi} \textsf{sin}\phi
    +0.5 k_f l_s^2 \textsf{sin}\phi \\
    +0.5 b_f l_s^2 \dot{\phi}\textsf{cos}\phi
    +0.5 k_r l_s^2 \textsf{sin}\phi +0.5 b_r l_s^2 \dot{\phi}\textsf{cos}\phi,
    \end{multline}
    where $\eta>0 $ is the sliding mode gain, $\psi > 0$ is a tunable hyperparameter, and all other symbols are defined in \cref{subsec:roll_model}.
\end{thm}

\begin{proof}
    We define a sliding surface $s \defeq \phi + \psi \dot{\phi}$ for the roll control model, where $\psi$ is a non-negative hyperparameter balancing stability (roll angle) and rider comfort (roll rate). The first time-derivative of the surface yields $\dot{s}=\dot{\phi} + \psi \ddot{\phi}$. We use the Lyapunov variant of sliding mode control to ensure stability, $\dot{s}=-\eta s$ to obtain the identity $\dot{\phi} + \psi \ddot{\phi} = -\eta s.$ Substituting $\cref{eq:roll}$ for $\ddot{\phi}$ into the identity, we get
    
    \begin{align*}
         &\Bigl(m_s a_y h_{\phi} \textsf{cos}\phi +m_s g h_{\phi} \textsf{sin}\phi
    -0.5 k_f l_s^2 \textsf{sin}\phi -0.5 b_f l_s^2 \dot{\phi}\textsf{cos}\phi
    -0.5 k_r l_s^2 \textsf{sin}\phi -0.5 b_r l_s^2 \dot{\phi}\textsf{cos}\phi
    -0.5 k_f l_s \left(z_{u,fl}-z_{u,fr}\right)\\ 
    &-0.5 b_f l_s \left(\dot{z}_{u,fl}-\dot{z}_{u,fr}\right)
    -0.5 k_r l_s \left(z_{u,rl}-z_{u,rr}\right) -0.5 b_r l_s \left(\dot{z}_{u,rl}-\dot{z}_{u,rr}\right)\Bigr) +  u_{\phi} = - \tilde{I}_{xx}\frac{\eta}{\psi} s -\tilde{I}_{xx} \frac{\dot{\phi}}{\psi}
    \end{align*}
    Rearranging the terms we obtain the roll control command 
    \begin{align*}
         u_{\phi} &= - \tilde{I}_{xx}\frac{\eta}{\psi} s -\tilde{I}_{xx} \frac{\dot{\phi}}{\psi} -m_s a_y h_{\phi} \textsf{cos}\phi -m_s g h_{\phi} \textsf{sin}\phi
    +0.5 k_f l_s^2 \textsf{sin}\phi +0.5 b_f l_s^2 \dot{\phi}\textsf{cos}\phi
    +0.5 k_r l_s^2 \textsf{sin}\phi\\ &+0.5 b_r l_s^2 \dot{\phi}\textsf{cos}\phi
    +0.5 k_f l_s \left(z_{u,fl}-z_{u,fr}\right)
    +0.5 b_f l_s \left(\dot{z}_{u,fl}-\dot{z}_{u,fr}\right)
    +0.5 k_r l_s \left(z_{u,rl}-z_{u,rr}\right) +0.5 b_r l_s \left(\dot{z}_{u,rl}-\dot{z}_{u,rr}\right)
    \end{align*}
    We have $0.5 k_f l_s \left(z_{u,fl}-z_{u,fr}\right) << 0.5 k_f l_s^2 \textsf{sin}\phi$ since $\lvert z_{u,fl}-z_{u,fr}\rvert <= 0.1 \;\textsf{m}$, considering a maximum allowable relative displacement of the suspension between the sprung and unsprung masses $\lvert z_{s}-z_{u,ij} \rvert <= 0.1\; \textsf{m} $ for all $i\in\{\textsf{front}\; (f), \textsf{rear}\; (r)\}$, $j\in\{\textsf{left}\; (l), \textsf{right}\; (r)\}$. Analogous results exist for $0.5 b_f l_s \left(\dot{z}_{u,fl}-\dot{z}_{u,fr}\right)$, $
    0.5 k_r l_s \left(z_{u,rl}-z_{u,rr}\right)$, and $0.5 b_r l_s \left(\dot{z}_{u,rl}-\dot{z}_{u,rr}\right)$ due to general mechanical restrictions imposed upon active suspension systems. Without loss of generality, we remove these terms from the roll control command to get
    \begin{align*}
         u_{\phi} &\approx - \tilde{I}_{xx}\frac{\eta}{\psi} s -\tilde{I}_{xx} \frac{\dot{\phi}}{\psi} -m_s a_y h_{\phi} \textsf{cos}\phi -m_s g h_{\phi} \textsf{sin}\phi
    +0.5 k_f l_s^2 \textsf{sin}\phi +0.5 b_f l_s^2 \dot{\phi}\textsf{cos}\phi
    +0.5 k_r l_s^2 \textsf{sin}\phi +0.5 b_r l_s^2 \dot{\phi}\textsf{cos}\phi\\
    \end{align*}
    Finally substituting $s$, 
    \begin{multline*}
        u_{\phi} \approx - \tilde{I}_{xx}\frac{\eta}{\psi} \phi - \tilde{I}_{xx}\left(\eta+\frac{1}{\psi}\right)\dot{\phi} -m_s a_y h_{\phi} \textsf{cos}\phi -m_s g h_{\phi} \textsf{sin}\phi
    +0.5 k_f l_s^2 \textsf{sin}\phi \\
    +0.5 b_f l_s^2 \dot{\phi}\textsf{cos}\phi
    +0.5 k_r l_s^2 \textsf{sin}\phi +0.5 b_r l_s^2 \dot{\phi}\textsf{cos}\phi
    \end{multline*}

\end{proof}
	We can reduce the aforementioned $u_{\phi}$ formulation further by making small angle approximations for computational simplicity. In the subsection below, we present a brief outline for the asymptotic stability and robustness of this controller.
\subsection{Stability and robustness}
It is trivial to establish Lyapunov (asymptotic) stability for the given sliding surface~\cite{polyakov2014stability}. Consider a Lyapunov function $V \left(\phi,\dot{\phi}\right) \defeq \frac{1}{2} s^2$. Taking the time derivative, we obtain
\begin{align*}
         \dot{V} &\left(\phi,\dot{\phi}\right) \defeq \frac{1}{2} \;2\; s\; \dot{s}\\
         \implies \dot{V} &\left(\phi,\dot{\phi}\right) \defeq -\eta s^2,\; \because \; \dot s = -\eta s \\
         \textsf{Therefore,}\; \dot{V} &\left(\phi,\dot{\phi}\right) \defeq -\eta s^2 < 0, \; \forall \; \eta > 0.
    \end{align*}

 Sliding mode controllers are inherently robust to uncertainties~\cite{erbatur1999study}. Further we can bound any modeling uncertainties by introducing a third tunable hyperparameter in conjunction with a $\textsf{sign}(s)$ to ensure robustness stability~\cite{chu2015smooth}.

\subsection{Implementation of the controller}\label{subsec:implementation}
The solution to \cref{thm:sliding_mode} generates an asymptotically stable, robust controller, which can be further simplified for implementation. The key measurements required for computation of $u_{\phi}$ include roll angle $\phi$, roll rate $\dot{\phi}$, and lateral acceleration $a_y$. While $\dot{\phi}$ is directly measured by a gyroscope placed at the center of gravity of the vehicle, $\phi$ is obtained via a derivative or Kalman filtering. The lateral acceleration measurement is often prone to errors and noise, and requires further signal processing prior to use. This process introduces unacceptable time-delays that could reduce the effectiveness of the roll control algorithm. Therefore we leverage the relationship between lateral acceleration and steering angle for a given longitudinal velocity $a_y \approx \frac{\delta_{\textsf{front}} \dot{x}^2}{l_{w}+K_u m_s \dot{x}^2}$~{\!\!\cite[Eq.~4.28]{jacobson2016vehicle}}, where $\delta_{\textsf{front}}$ is the steering angle, $l_w$ is the length of the wheel base, $\dot{x}$ is longitudinal velocity, and $K_u$ is the vehicle understeer gradient, to obtain
\begin{multline}
        u_{\phi} = - \tilde{I}_{xx}\frac{\eta}{\psi} \phi - \tilde{I}_{xx}\left(\eta+\frac{1}{\psi}\right)\dot{\phi} - m_s\frac{\delta_{\textsf{front}} \dot{x}^2}{l_{w}+K_u m_s \dot{x}^2} h_{\phi} \textsf{cos}\phi -m_s g h_{\phi} \textsf{sin}\phi
    +0.5 k_f l_s^2 \textsf{sin}\phi \\
    +0.5 b_f l_s^2 \dot{\phi}\textsf{cos}\phi
    +0.5 k_r l_s^2 \textsf{sin}\phi +0.5 b_r l_s^2 \dot{\phi}\textsf{cos}\phi
    \end{multline}
    
\noindent While the roll control command is computed in the Vehicle controls unit (VCU)/ electronic controls unit (ECU), the actuation takes place via the four linear motors connected to the suspension at the four corners of the vehicles. Distribution of heave control, pitch control, and roll control commands to the four corners occurs via the relationship:
\begin{equation}
    \left[\begin{smallmatrix}
        u_{z}\\
        u_{\theta}\\
        u_{\phi}
    \end{smallmatrix}\right]= \left[\begin{smallmatrix}
        -1 & -1 & -1 & -1\\
        a & a & a-l & a-l\\
        \frac{l_s}{2} & -\frac{l_s}{2} & \frac{l_s}{2}& -\frac{l_s}{2}
    \end{smallmatrix}\right]\left[\begin{smallmatrix}
        F_{fl}\\
        F_{fr}\\
        F_{rl}\\
        F_{rr}
    \end{smallmatrix}\right],
\end{equation}
where $u_z$ is the heave control command, $u_{\theta}$ is pitch control, $a$ is the length of the vehicle from the front to center of gravity, $l$ is full length of the vehicle, $F_{fl}$, $F_{fr}$, $F_{rl}$, $F_{rr}$ represent the vertical force commands at the front-left, front-right, rear-left, and rear-right wheels respectively. All other symbols have been defined previously in \cref{subsec:roll_model}. Define the gain matrix as $A$ for simplicity in explanation. We determine the vertical forces at the wheels by taking the pseudo-inverse of A and multiplying with the control commands vector. For $u_{z} = u_{\theta} = 0$, the problem reduces to $F = \left(A^\tp A\right)^{-1}\left(\begin{smallmatrix}
    \frac{l_s}{2} & -\frac{l_s}{2} & \frac{l_s}{2}& -\frac{l_s}{2}
\end{smallmatrix}\right)^\tp u_{\phi}$.  
\subsection{Effect of road bank angle}\label{subsec:road_bank}
The road geometry plays a critical role in vehicle handling scenarios. Among such road parameters, road bank angle plays a critical role in mitigation of vehicle rollover situations. However this cannot be directly measured by low cost sensors. Several attempts have been made to estimate the road bank angle separately from the roll angle using observed-based techniques~\cite{ryu2004estimation,grip2009estimation,tseng2001dynamic}. The estimate of the roll bank angle ($\phi_{\textsf{road}}$) can be incorporated for better roll control by modifying the sliding surface $s \defeq \phi-\phi_{\textsf{road}} + \psi \dot{\phi}$. Thus the roll control signal can be modified as
\begin{multline}
        u_{\phi} = - \tilde{I}_{xx}\frac{\eta}{\psi} \left(\phi-\phi_{\textsf{road}}\right) - \tilde{I}_{xx}\left(\eta+\frac{1}{\psi}\right)\dot{\phi}- \tilde{I}_{xx}\frac{1}{\psi}\dot{\phi}_{\textsf{road}} - m_s \tilde{a}_y h_{\phi} \textsf{cos}\phi -m_s g h_{\phi} \textsf{sin}\left(\phi-\phi_{\textsf{road}}\right)\\
    +0.5 k_f l_s^2 \textsf{sin}\phi 
    +0.5 b_f l_s^2 \dot{\phi}\textsf{cos}\phi
    +0.5 k_r l_s^2 \textsf{sin}\phi +0.5 b_r l_s^2 \dot{\phi}\textsf{cos}\phi,
    \end{multline}
where $\tilde{a}_y = a_y -g \;\textsf{sin}\phi_{\textsf{road}} \approx \frac{\delta_{\textsf{front}} \dot{x}^2}{l_{w}+K_u m_s \dot{x}^2} -g \;\textsf{sin}\phi_{\textsf{road}}.$
\section{Experimental Results}\label{sec:experiments}
In this section, we experimentally validate the roll control algorithm on an electric car, equipped with linear motors for vertical translations at the four corners of the vehicle. We perform MATLAB car simulations using the following parameters for the vehicle: sprung mass $m_s = 820 $kg, unsprung mass including tire $m_u = 60$ kg, sprung roll inertia moment $I_{xx}=120$ kg-$m^2$, roll center $h_{\phi}=0.48$ m, wheelbase of $2.3$m, front and rear track-widths of $1.3$m and $1.3$m, front and rear spring constants of $12000$ N/m and $35000$ N/m respectively,  front and rear damping constants of $530$ N-s/m and $850$ N-s/m respectively.
\subsection{Slalom Test}
\begin{figure}[hbt!]
	\centering
 	\includegraphics[width=0.6\linewidth]{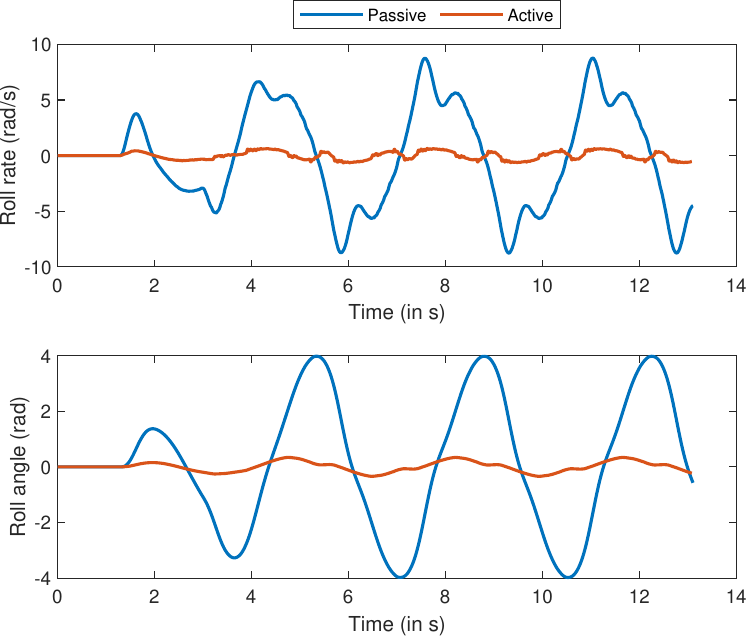}
	  \caption{Plot of the roll angle and the roll rate for an active suspension electric vehicle undergoing the slalom test at a constant longitudinal velocity of 30 kph for approximately 13 s. There is more than $90\%$ reduction in the peak roll rate for active suspension roll control in comparison to passive.}
	\label{fig:Slalom_30kph}
\end{figure}

\begin{figure}[hbt!]
	\centering
 	\includegraphics[width=0.6\linewidth]{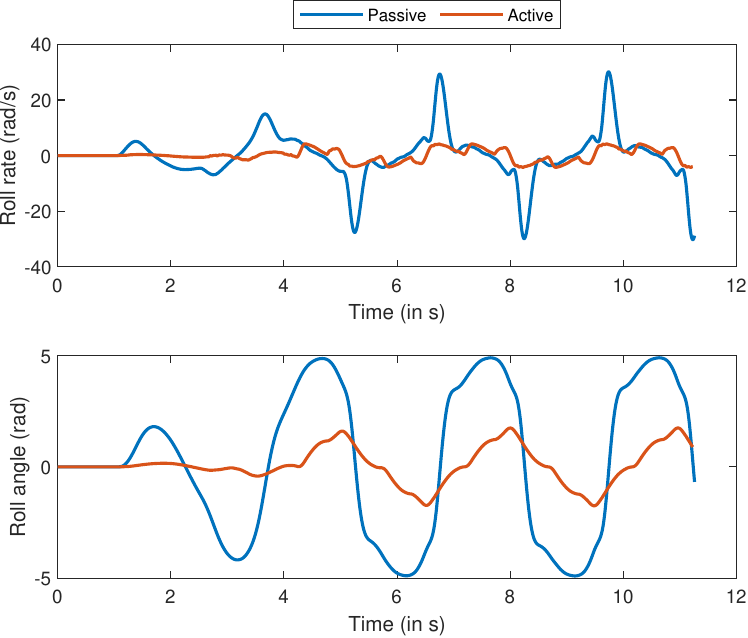}
	  \caption{Plot of the roll angle and the roll rate for an active suspension electric vehicle undergoing the slalom test at a constant longitudinal velocity of 35 kph for approximately 13 s. There is a continued $90\%$ reduction in the peak roll rate for active suspension roll control in comparison to passive. All conditions remain the same as for the 30 kph test.}
	\label{fig:Slalom_35kph}
\end{figure}

\begin{figure}[hbt!]
	\centering
 	\includegraphics[width=0.6\linewidth]{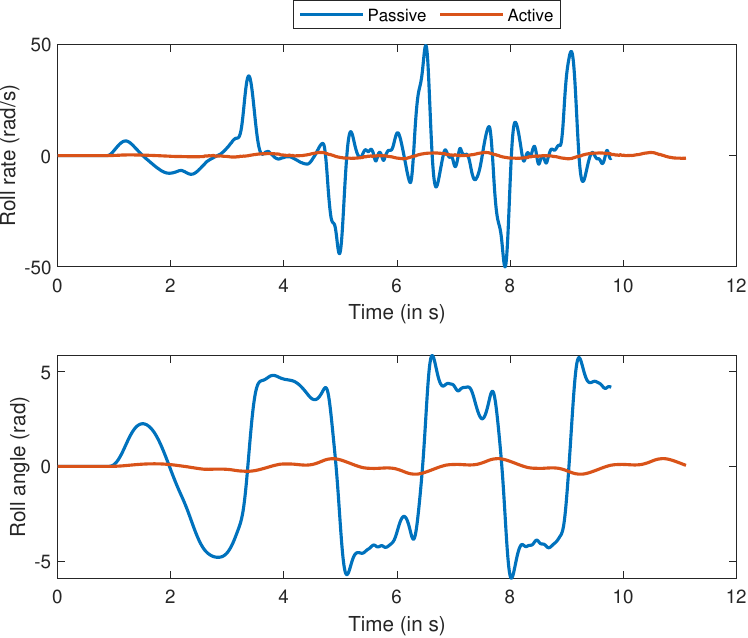}
	  \caption{Plot of the roll angle and the roll rate for an active suspension electric vehicle undergoing the slalom test at a constant longitudinal velocity of 40 kph for approximately 13 s. There is a continued $90\%$ reduction in the peak roll rate for active suspension roll control in comparison to passive. All conditions remain the same as for the 30 kph test.}
	\label{fig:Slalom40kph}
\end{figure}

Slalom tests requires a vehicle to be driven in a straight line, followed by a sequence of turns around cones in alternate directions with constant vehicle velocity. This test simulates vehicle maneuvers during cornering or turning or during avoidance of sudden obstacles. The vehicle is driven in a straight line at a velocity slightly higher than the desired longitudinal velocity, following which the driver releases the throttle, coasting to the test velocity entering a winding course of traffic cones spaced at intervals of $15.24$ m at a constant speed. We simulate this test in MATLAB at 30 kph, 35 kph, and 40 kph. \cref{fig:Slalom_30kph,fig:Slalom_35kph,fig:Slalom40kph} represent the behavior of the roll angle and roll rate for a passive system (in blue) and active system with sliding mode roll control switched on (in red). It can be clearly observed that the roll angle has been reduced consistently by atleast $60\%$ compared to the passive, ensuring vehicle stability. We can observe the increase in spikes in the roll rate as longitudinal speed increases in comparison to with active roll control. This translates into significant `jerk' experienced by the driver during cornering events, which is virtually reduced with sliding mode roll control.
\label{subsec:slalom}

\begin{figure}[hbt!]
	\centering
 	\includegraphics[width=0.6\linewidth]{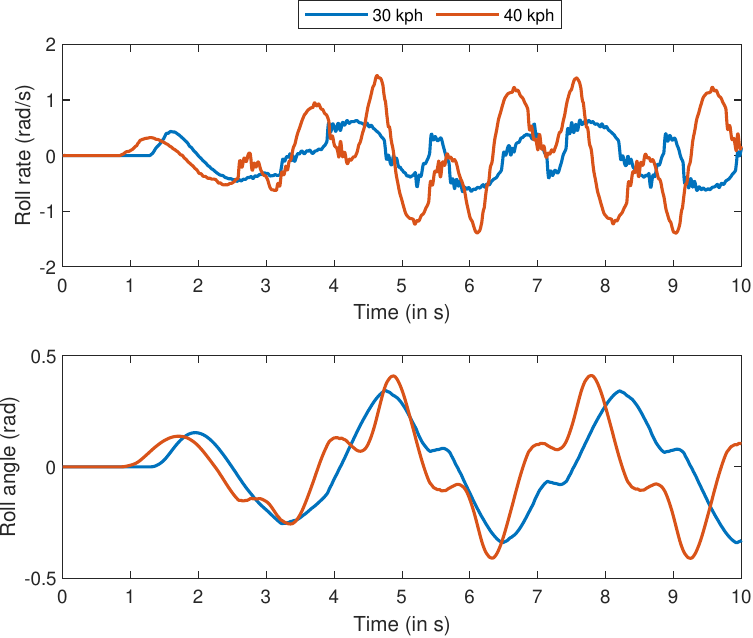}
	  \caption{A closeup snapshot of the roll angle and roll rate for active suspension based sliding model roll control. We can notice the increase in perturbations (peakiness) as we increase the constant longitudinal speed from 30 kph to 40 kph. However as observed from the plots in \cref{fig:Slalom_30kph,fig:Slalom40kph} the active roll control is substantially better than passive suspension systems.}
	\label{fig:ARC_30_40}
\end{figure}

\subsection{Role of Driver Preview}
\begin{figure}[hbt!]
	\centering
 	\includegraphics[width=0.6\linewidth]{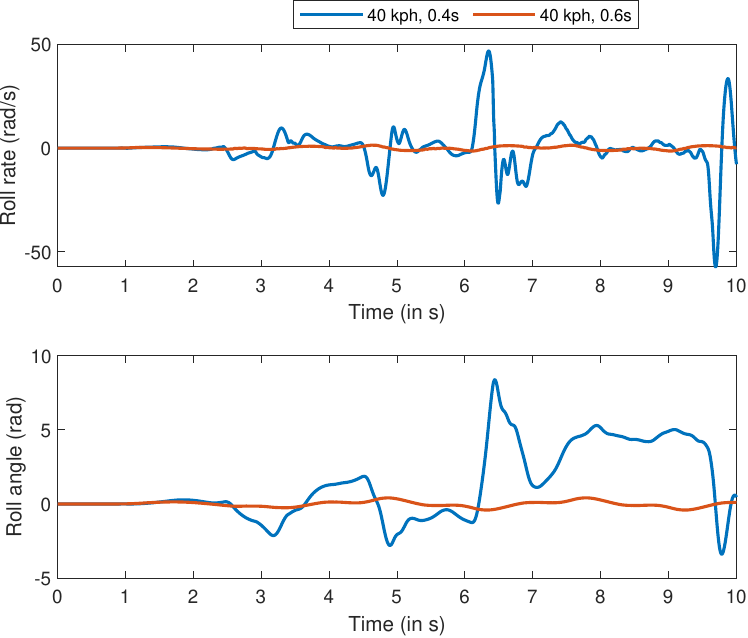}
	  \caption{Plot of the roll angle and the roll rate for an active suspension electric vehicle undergoing the slalom test at a constant longitudinal velocity of 40 kph for approximately 10 s. There is a continued $90\%$ reduction in the peak roll rate for active suspension roll control with a preview time of 0.6 seconds compared to 0.4 seconds.}.
	\label{fig:Preview_ARC_40}
\end{figure}

\begin{figure}[hbt!]
	\centering
 	\includegraphics[width=0.6\linewidth]{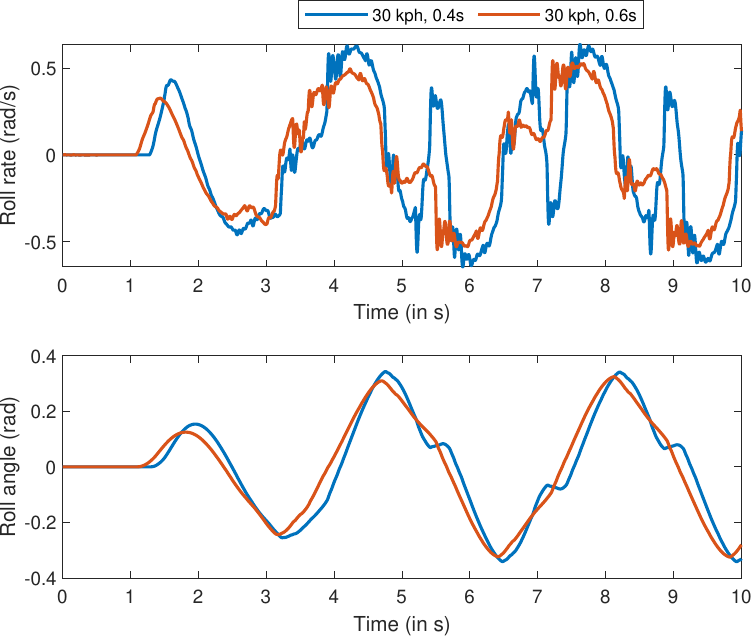}
	  \caption{Plot of the roll angle and the roll rate for an active suspension electric vehicle undergoing the slalom test at a constant longitudinal velocity of 30 kph for approximately 10 s. There is not a significant difference in peak roll rate and roll angle with preview time of 0.6 seconds compared to 0.4 seconds.}.
	\label{fig:Preview_ARC_30}
\end{figure}

Driver preview time has significant influence on driving behavior, especially in situations like turning and cornering \cite{zhou2021vehicle}. Based on multiple studies, it has been established that the preview time is generally longer than required time for the driver's information
process, and the minimum preview time to keep the vehicle stable was around 3 seconds \cite{kono2011review}. We simulate Slalom tests for two preview times of 0.4 seconds and 0.6 seconds for 30 kph in \cref{fig:Preview_ARC_30} and 40 kph in \cref{fig:Preview_ARC_40}. While a higher preview time of 0.6 seconds marginally improves the roll rate by reducing the spikiness of the curve for a velocity of 30 kph, it significantly reduces the roll rate and roll angle for a vehicle velocity of 40 kph and ensures more smoother curves for roll and roll rate. These experiments establish the effectiveness of the proposed control method at extremely low values of driver preview time.
\label{subsec:preview}

\subsection{J-Turn Test}
\begin{figure}[hbt!]
	\centering
 	\includegraphics[width=0.6\linewidth]{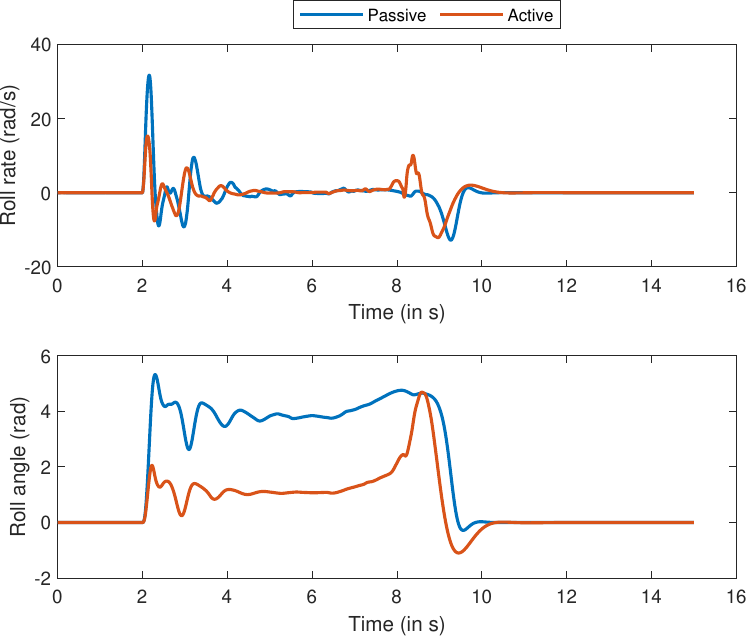}
	  \caption{Plot of the roll angle and the roll rate for an active suspension electric vehicle undergoing the J-turn maneuver at a constant longitudinal velocity of 60 kph for approximately 14 s. There is more than $45\%$ reduction in the peak roll rate for active suspension roll control in comparison to passive.}.
	\label{fig:Jturn_60}
\end{figure}

\begin{figure}[hbt!]
	\centering
 	\includegraphics[width=0.6\linewidth]{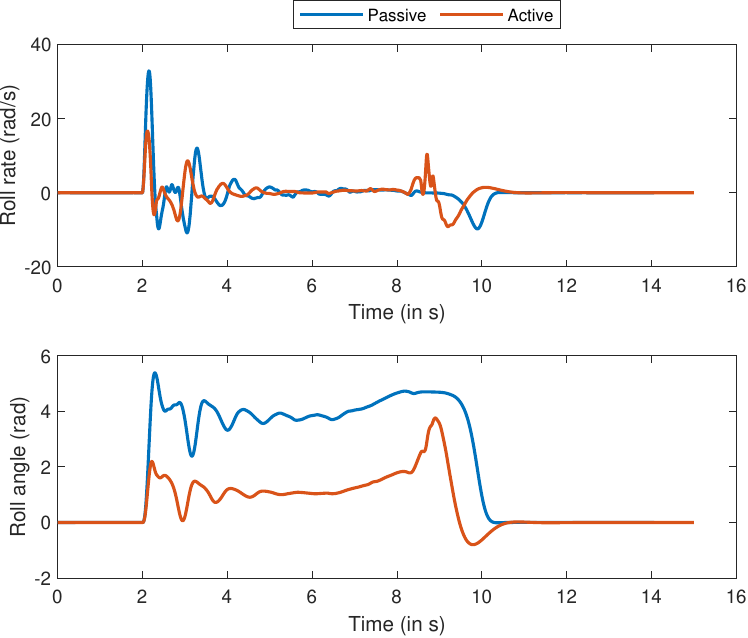}
	  \caption{Plot of the roll angle and the roll rate for an active suspension electric vehicle undergoing the J-turn maneuver at a constant longitudinal velocity of 72 kph for approximately 14 s. There is more than $45\%$ reduction in the peak roll rate for active suspension roll control in comparison to passive. Test conditions remain the same as for the 60 kph test.}
	\label{fig:Jturn_72}
\end{figure}

\begin{figure}[hbt!]
	\centering
 	\includegraphics[width=0.6\linewidth]{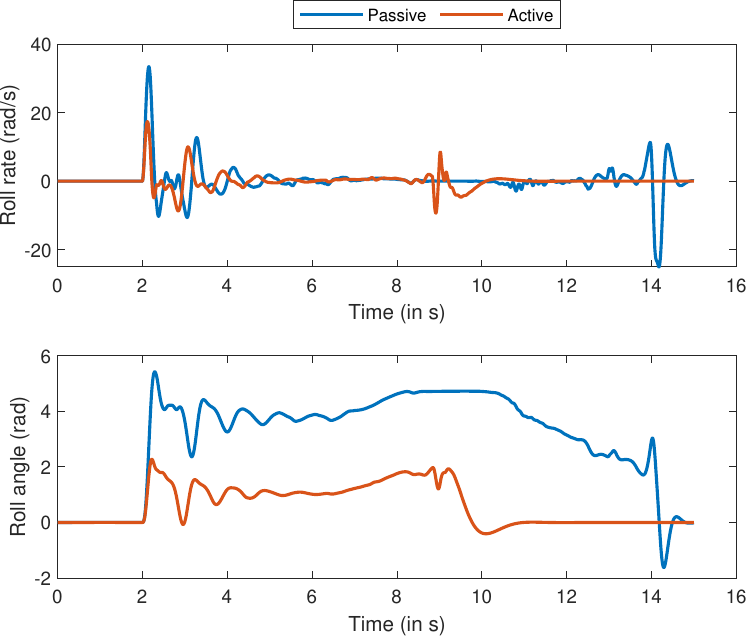}
	  \caption{Plot of the roll angle and the roll rate for an active suspension electric vehicle undergoing the J-turn maneuver at a constant longitudinal velocity of 80 kph for approximately 14 s. There is more than $45\%$ reduction in the peak roll rate for active suspension roll control in comparison to passive. Test conditions remain the same as for the 60 kph test. However, passive suspension control is substantially delayed as velocity increases, which is compensated by active roll control because of its faster response.}
	\label{fig:Jturn_80}
\end{figure}

The vehicle is simulated performing an NHTSA standardized J-turn test that represents an avoidance maneuver in which the vehicle is steered away from an obstacle using a ramp steering input~\cite{forkenbrock2005nhtsa,zhang2008investigation}. The test conditions requires the handwheel angle producing a steady-state lateral acceleration of 0.3 g at 50 mph on a level paved surface. The vehicle is driven in a straight line at a velocity slightly higher than the desired longitudinal velocity, following which the driver releases the throttle, coasting to the test velocity, and then triggers the commanded handwheel input. The nominal longitudinal velocity for this test is from 35 to 60 mph, increased in 5 mph increments until termination conditions are satisfied. Termination conditions are simultaneous two inch or greater lift of the vehicle's inside tires or when the test is completed at maximum possible test longitudinal velocity without two wheels lift~{\!\!\cite[Ch.~15]{rajamani2011vehicle}}. We simulate this test in MATLAB at 60 kph ($35$ mph), 72 kph ($45$ mph), and 80 kph ($50$ mph). \cref{fig:Jturn_60,fig:Jturn_72,fig:Jturn_80} represent the behavior of the roll angle and roll rate for a passive system (in blue) and active system with sliding mode roll control switched on (in red). It can be clearly observed that the roll angle has been reduced consistently by atleast $50\%$ with sliding model active suspension based roll control compared to a passive suspension vehicle, ensuring vehicle stability. We can also note that the active control has faster response time compared to the passive case. Further the roll rate is reduced significantly ($> 45\%$) peak to peak compared to passive in \cref{fig:Jturn_60}, and this value remains approximately the same with increase in longitudinal velocity.  
\subsection{Impact of $\eta$}
\begin{figure}[ht]
	\centering
 	\includegraphics[width=0.6\linewidth]{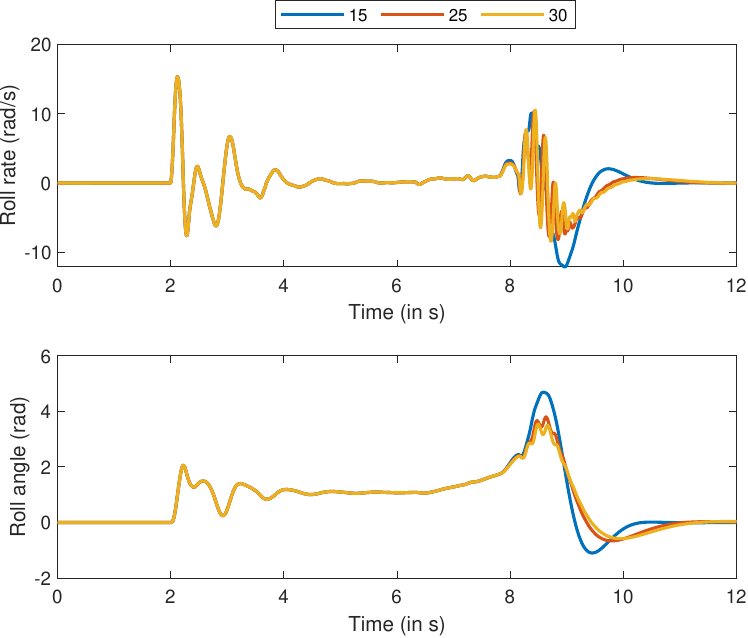}
	  \caption{J-turn maneuver at a longitudinal velocity of 60 kph for sliding gains 15, 25, and 30. Higher sliding gains introduce oscillations in the roll rate leading to a bumpier ride.}
	\label{fig:Jturn_eta_60}
\end{figure}

\begin{figure}[ht]
	\centering
 	\includegraphics[width=0.6\linewidth]{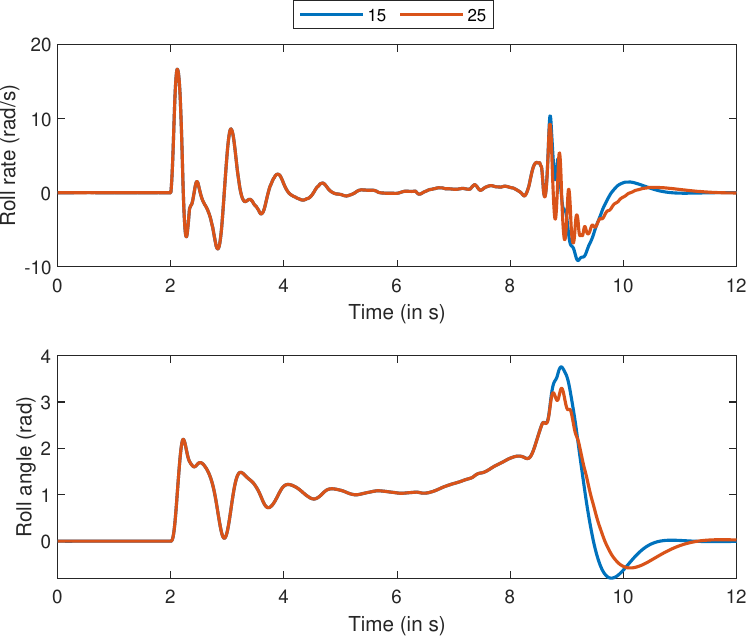}
	  \caption{J-turn maneuver at a longitudinal velocity of 72 kph for sliding gains 15,  and 25. Higher sliding gains introduce oscillations in the roll rate leading to a bumpier ride and produce a delayed control response.}
	\label{fig:Jturn_eta_72}
\end{figure}

In this experiment, we study the variable impact of the sliding gain $\eta$ during a J-turn maneuver for vehicle velocities of 60 kph in \cref{fig:Jturn_eta_60} and 72 kph in \cref{fig:Jturn_eta_72}. It is conclusive from the experiments that a lower sliding gain of 15 performs better compared to a higher value. As we increase the $\eta$, the control response is delayed during the later phase of the J-turn maneuver, with substantial oscillations introduced in the roll rate curve, indicating a bumpier ride.  
\section{Conclusions}
\label{sec:conclufuture}
We considered a novel roll control algorithm based on the well known sliding mode control technique for mitigating vehicle rollovers in electric vehicles equipped with active suspensions. Specifically, we leveraged a 6 DoF roll dynamics model by combining a sprung mass roll dynamics model with a quarter car model for the unsprung masses to design this roll controller. We provide a practical implementation of this controller by leveraging the relationship between lateral acceleration and vehicle steering angle to obtain a faster, simplified roll control implementation, which is distributed across the four vehicle corners equipped with electric actuators capable of generating vertical force commands. We validate the effectiveness and functionality of the roll control algorithm via numerical simulations. Our study shows an approximately 50$\%$ reduction in the propensity of rollover by reducing the roll angle, while ensuring a smoother ride by a 90$\%$ reduction in the roll rate.
\newpage
\if\MODE3
\bibliographystyle{IEEEtran}
\bibliography{rdlqr}
\else
\bibliographystyle{abbrv}
{\small \bibliography{rdlqr}}
\fi

\end{document}